# Optimization of Beam Transmission in the extraction lines of the K130 Room Temperature Cyclotron at Variable Energy Cyclotron Centre (VECC), Kolkata

Arup Ratan Jana[*], Animesh Goswami and Malay Kanti Dey

*Variable Energy Cyclotron Centre, 1/AF, Bidhannagar, Kolkata-700064, India*

*Homi Bhabha National Institute, Anushaktinagar, Mumbai - 400 094, India*

**Abstract**

The K130 Room Temperature Cyclotron (RTC) at the Variable Energy Cyclotron Centre (VECC), Kolkata has delivered a variety of ion beams for experimental programs with stable performance since the 1970s. The extracted beams are transported to experimental stations through four beamlines. In this work, a two-step semi-numerical optimization framework is presented to improve the transmission efficiency of these transport lines. At its core, the method relies on multiparticle simulations to account for beam losses and nonlinear effects, which render the problem analytically intractable. To address these complexities, a derivative-free global optimization framework is developed to systematically improve beam transmission in the cyclotron beamlines, and its effectiveness is demonstrated experimentally.

In the first phase, the beam parameters at the beamline entrance are estimated using Bayesian optimization constrained by experimentally measured transmission ratios for selected optics configurations. In the second phase, the quadrupole gradients are optimized using the Covariance Matrix Adaptation Evolution Strategy (CMA-ES) to maximize beam transmission. Optimized settings computed using this simulation based semi numerical methodology were then experimentally validated, achieving beam transmissions exceeding 70%.

[*] Corresponding author. *E-mail* : ar.jana@vecc.gov.in

# Contents

## 1. Introduction

The Room Temperature Cyclotron (RTC) at the Variable Energy Cyclotron Centre (VECC), Kolkata, has been in continuous operation for almost five decades [1]. It is a K130 cyclotron, where the K-value defines the maximum achievable kinetic energy of ions as a function of their charge to mass ratio [2-4]. The K130 cyclotron is a three-sector machine equipped with a water-cooled 262 tons magnet of pole diameter of 224 cm, producing a maximum average magnetic field of 17 kG. Particles in the cyclotron are accelerated by a radio-frequency (RF) system comprising a 180° Dee and a strip-lined dummy Dee. Operating in the frequency range of 5.5–8.5 MHz this RF system provides a maximum Dee voltage of about 65 kV [5]. The machine is currently equipped with an internal PIG ion source, which routinely delivers light-ion beams such as protons, deuterons and alpha particles for various experimental use. In addition, an indigenously developed 14.45 GHz external ECR ion source is also available for heavy-ion beams including nitrogen, oxygen, neon and argon in multiple charge states, with beam currents ranging from a few picoamperes to several microamperes [6]. Finally, the beam is extracted from the cyclotron using an electrostatic deflector at the desired energy. Then the extracted beam enters a transport section that is common to all beamlines. From there, downstream of this section, a switching magnet is employed to direct the beam into the desired experimental beamline by adjusting its field strength.

The beam is delivered to user facilities through four dedicated transport channels, referred to as RTCBL-01 to RTCBL-04. Each beamline supports different experimental requirements by providing ions of specific species and energies. The beam lines RTCBL-01, RTCBL-02 and RTCBL-03 are oriented at 0°, 20°12′ and 35°12′ respectively, relative to the beam extraction axis. Beamline RTCBL-01 is utilized for material science, radiation damage studies, radiochemistry and isotope production whereas RTCBL-02 and RTCBL-03 are primarily employed for nuclear physics research. RTCBL-01 and RTCBL-04 share a common initial section. When the 160° analyzing magnet is energized, it directs the beam into RTCBL-04 beam line. This beam line RTCBL-04 is dedicated to the production and investigation of radioactive ion beams (RIBs) [7]. Among these four beamlines, RTCBL-01, RTCBL-02 and RTCBL-03 are equipped with six quadrupole and six steering magnets to maintain beam quality and alignment. RTCBL-04 includes two additional steering magnets to support the operational needs of RIB production. The beam envelope in these transport lines is tuned primarily by adjusting the magnetic strengths of the focusing and defocusing quadrupole magnets. There, the steering magnets are used as required for orbit centring and alignment during routine operation. The overall layout and schematics of the beam transport system are shown in Fig. 1[8].

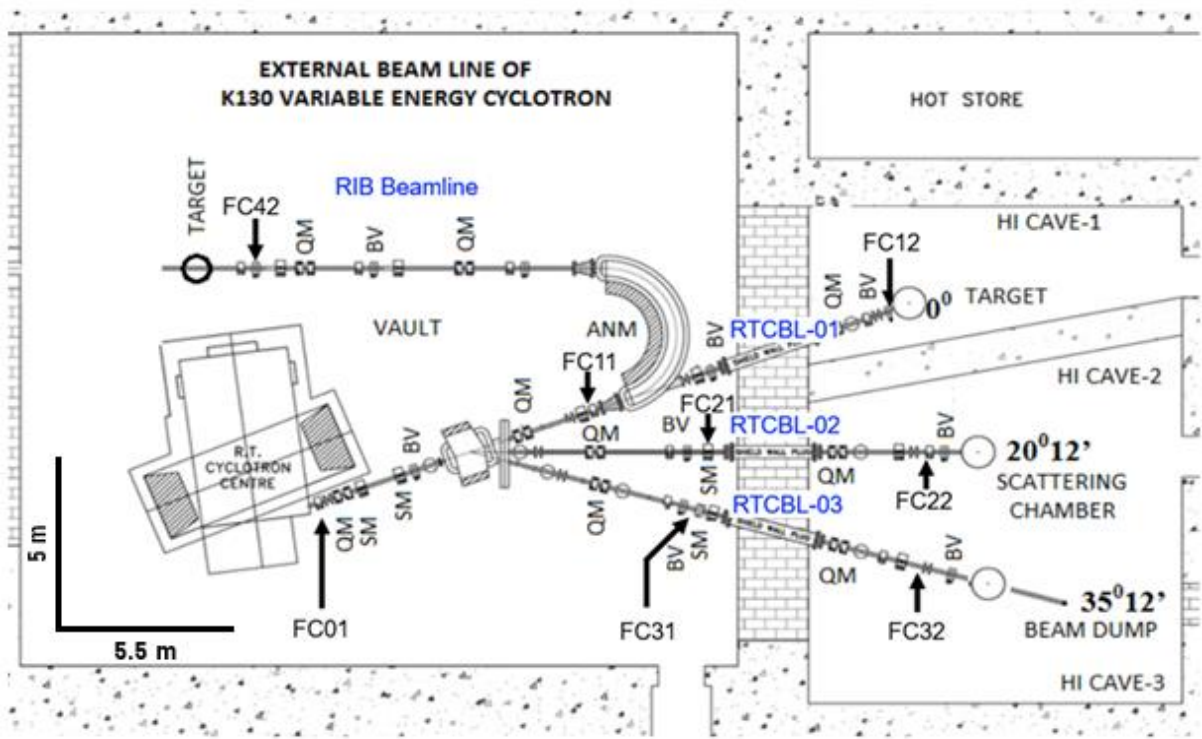

**Fig. 1:** Layout of a section of the four beamlines (top) and their corresponding schematic representations (bottom). QM, SM, and ANM represent Quadrupole Doublets, Steering Magnets, and the Analyzing Magnet, respectively. FC and BV indicate the Faraday Cup and Beam Viewer, respectively.

In order to meet the users demand for high-current, light ion operation using PIG ion source, it requires careful tuning of the ion sources operating in high-yield mode. This increases beam output but shortens filament lifetime and reduces long-term reliability. Apart from that, the acceleration of intense beams within the cyclotron further leads to increased particle losses, contributing to activation of internal components. Since the present extraction efficiency of the cyclotron is limited to approximately 20%, these losses significantly enhance residual radioactivity and complicate long-term operation as well as hands-on maintenances. A more suitable approach to achieve high target current therefore lies in improving beam transport efficiency through optimizing the beam lines rather than simply increasing source current. The present study therefore focuses on improving beam transmission in the transport line through systematic optimization of the quadrupole magnets. A simulation-based semi-numerical optimization framework is developed for this purpose. In these transport lines, the steering magnets are used only for orbit centring during routine operation and are therefore not included in the present optimization framework. The proper adjustment of these focusing elements is crucial for controlling the beam envelope and trajectory along the transport channel. However, reliable optimization essentially requires precise knowledge of the input beam conditions, particularly the beam emittances and Twiss parameters at the entrance of the beam line. It is therefore, essential to determine these parameters at the entrance of the transport line and use them as the starting point for beam dynamics simulations to predict and improve transmission efficiency.

To the best of our knowledge, the parameters of the beam extracted from the K130 cyclotron have not yet been fully established through detailed beam dynamics analysis or systematic experimental measurements. Moreover, implementing routine measurements of the beam parameters at the entrance of these beamlines in the RTC poses unique challenges. The compact and decades-old design offers little scope for installing additional diagnostics. Furthermore, the cyclotron operates almost continuously to meet the demands of a large user community [9] [10], both national and international, making extended measurement campaigns impractical. Therefore, in practice, for the K130 room-temperature cyclotron, lattice settings have historically evolved through operational experience. Adjustments have been made incrementally over years of routine operation to satisfy user requirements. While this empirical approach has confirmed reliable day-to-day operation, it lacks analytical capability and limits the scope for systematic improvement.

Given the practical constraints and limitations, a computationally based, reliable optimization framework becomes essential for the RTC beamlines. To meet this requirement, we have developed machine-specific semi-numerical optimization methodology for beamline optimization. The method combines a Python-based framework with the standard beam dynamics code TRACEWIN [11-12]. It has been benchmarked and validated and the present work reports its development and application. This approach allows for a systematic and computationally efficient optimization without requiring additional hardware or interrupting machine operation.

In this developed framework, the beam dynamics code TRACEWIN plays an important role. Therefore, it is useful to briefly discuss a few relevant features of this code as employed in the present work. TRACEWIN is a versatile beam dynamics simulation tool that supports two principal calculation modes: an *rms* beam envelope mode and a multiparticle tracking mode. The envelope mode is based on linear beam optics and propagates the *rms* beam parameters through the lattice using envelope equations. In contrast, the multiparticle mode performs numerical tracking of macroparticles through the beamline elements and naturally includes nonlinear effects arising from realistic electromagnetic fields, finite beam pipe apertures, and particle losses. Space-charge interactions between macroparticles can also be included when required. In the present analysis, TRACEWIN-based multiparticle beam dynamics simulations are performed, including space-charge forces to account for the finite beam current. Beam loss arises primarily from the finite aperture of the beam pipe. However, the space-charge effect remains relatively small due to the low beam current in the RTC beamlines, whereas aperture-induced beam loss plays a more significant role in the beam dynamics along the transport line. The beam transmission ratio at any location along the transport line is calculated from the fraction of macroparticles that survive up to that position. Accordingly, the total number of macroparticles used in the simulation is chosen to be fifty thousand, providing a reasonable compromise between computational cost and the statistical accuracy of the calculated transmission ratio.

The paper is organized as follows. Section 2 discusses the challenges associated with beamline optimization in the RTC and outlines the theoretical background of the problem. It also describes the semi-numerical methodology and its implementation. Section 3 presents the results of the analysis, including comparisons with operational data. Finally, Section 4 summarizes the conclusions and highlights possible directions for future work.

## 2. Theoretical Background and Numerical Methodology

In this section, we outline the methodology adopted for optimizing the beam transmission in the RTC beamlines. The first part of this methodology involves determining the unknown input beam parameters at the entrance of the beamline using a Python-based Bayesian optimization framework coupled with the standard multiparticle beam dynamics code TRACEWIN [13]. In the second part, the gradients of the six quadrupole magnets are optimized using CMA-ES based algorithm to achieve maximum beam transmission [14]. Here, multiparticle simulations are performed to account for beam losses and nonlinear effects. The subsequent subsections provide a detailed explanation of this two-step procedure.

*2.1 Problem Description*

As mentioned earlier, a semi-numerical optimization methodology is developed to enhance beam transmission through RTC beamlines. The framework of this semi-numerical methodology is implemented using a Python-based optimization framework integrated with the multiparticle beam dynamics code TRACEWIN, executed as a sub-process to perform simulations. Before moving ahead, we provide a brief overview of the focusing and correction elements in the four beamlines of the K130 cyclotron. As stated earlier, the lattice of each beamline consists of six quadrupole magnets that provide the required focusing to the beam. Along with these quadrupole magnets, the beamlines, RTCBL-01, RTCBL-02 and RTCBL-03 each include six steering magnets to correct minor deviations of the beam deviated from the ideal beam trajectory. Compared to these three beamlines, the RIB beamline RTCBL-04 includes two additional steering magnets. Although these four beamlines consist of both quadrupoles and steering magnets, the present optimization study focuses solely on the six quadrupole magnets. Among these four beamlines, RTCBL-01 is the most demanding one, where applications in material science studies, radiochemistry and isotope production demand considerably high beam current. In contrast, RTCBL-04, the feeder line for the RIB facility, is the longest of the four beamlines, featuring an analyzing magnet with a bending angle of 160°.

Accordingly, these two beamlines were initially chosen for lattice optimization to enhance performance through improved beam transmission.

Beam transmission through the finite aperture of a beamline is strongly affected by beam orbit offsets within the focusing elements. In routine operation of the RTC beamlines, the beam trajectory is therefore first centred using steering magnets with the aid of beam viewers. Once a near-centred orbit is established, the subsequent evolution of the beam envelope and the associated aperture losses are governed primarily by the quadrupole focusing strengths rather than by residual orbit offsets.

In the present framework, orbit correction is treated as a prerequisite operational step. The subsequent numerical optimization focuses on the quadrupole gradients that determine the envelope dynamics and transmission efficiency of the beamline. Although steering magnets could, in principle, be included as additional optimization variables, doing so would require reliable orbit-related observables. Moreover, their inclusion would introduce strong coupling with the quadrupole settings. This would make it difficult to isolate their individual effects, especially when the objective function is based solely on transmission measurements.

Given the limited diagnostics available for routine operation in this legacy machine, orbit correction is therefore treated separately from envelope optimization in the present implementation. A combined optimization of orbit and optics would be an important extension of this work once suitable orbit or position observables become available.

Based on the above discussion, it appears that the problem reduces to six variables corresponding to six quadrupole gradients. However, tracking the particle beam through the transport line also requires knowledge of the input beam parameters at the entrance as mentioned earlier. Since these input conditions are not available from prior measurements or calculations, they are evaluated computationally within the proposed methodology. The dimensionality of this optimization problem therefore increases from six to twelve variables, incorporating the emittances and the Twiss alpha ($\alpha$) and beta ($\beta$) parameters in each plane, along with the six quadrupole gradient values [15].

We have formulated the framework of this optimization problem as a two-part process: first, the determination of the unknown beam parameters, and second, the optimization of the beamline optics using these parameters.

*2.2 Estimation of input beam conditions*

In the first part of the problem, the unknown input beam parameters, including the horizontal and vertical beam emittances ($\varepsilon_x$, $\varepsilon_y$), together with the Twiss parameters $\alpha$ and $\beta$ in both transverse planes are evaluated. These parameters constitute the six variables that define the population in the evolutionary algorithm based optimization framework. The continuous (dc) input beam at the entrance of the beamline is assumed to be the same as the beam at the cyclotron exit. The particle distribution at the cyclotron exit is approximated by a four-dimensional Gaussian profile in the transverse phase space ($x$, $y$, $p_x$, $p_y$), described by these beam parameters, consistent with the Gaussian-like beam cores observed in measured transverse beam profiles reported in the literature [16-17]. As will be discussed later in this paper, the beam that actually propagates through the finite aperture of the beamlines corresponds to the truncated portion of this initial Gaussian distribution.

The lower and upper bounds assigned to these six variables in the two transverse plane, namely $\varepsilon_x$, $\varepsilon_y$, $\beta_x$, $\beta_y$, $\alpha_x$, $\alpha_y$ are summarized in Table 1. The size of population plays a significant role in determining the accuracy of the outcome and the speed of convergence. Based on this consideration, an initial population of 120 individuals corresponding to 20 times the number of variables, is generated within these specified ranges. In this calculation, lattice configurations along with the beam energy are fixed quantities. Later in this paper, we describe how these fixed lattices were chosen from the set of working beamline configurations used by the operators.

**Table 1:** The range of six variables at the beamline entrance.

| Parameter | Unit | Lower Limit | Upper Limit |
|---|---|---|---|
| $\varepsilon_x$, $\varepsilon_y$ | (mm-mrad) | 0.05 | 50 |
| $\beta_x$, $\beta_y$ | (mm/mrad) | 0.05 | 50 |
| $\alpha_x$, $\alpha_y$ | - | -50 | 0.0 |

The beam current is measured at several positions along the beamlines using Faraday Cups (FCs), which are indicated in the schematic of the K130 shown in Fig. 1. Based on these measurements, datasets of transmission ratios are obtained at two FC positions for several lattice configurations in a given beamline at a fixed beam energy, serving as inputs for constructing the reference benchmark. Here, we define the transmission ratio at any Faraday cup as the measured beam current at that location relative to FC-01 (faraday cup just at the cyclotron exit). In this study, the reference datasets were generated at the locations of FC-11 and FC-12 for the RTCBL-01 and for the RTCBL-04, they were obtained at FC-11 and FC-42. The developed optimization module iteratively evolves the population of input parameters toward convergence for each beamline. In this analysis, beam viewers (BV) are assumed to be used during operation for qualitative orbit centring and beam-spot inspection. Quantitative comparisons of the *rms* beam size are not reported here because the installed viewers were not calibrated for obtaining reproducible size across all operating conditions used in this study.

As mentioned earlier, the absence of reliable input beam parameters at the entrance and the occurrence of beam losses within the transport line make the problem analytically intractable. Therefore, within the kernel of the developed module, multiparticle tracking is carried out using TRACEWIN with fifty thousand particles in each run. These calculations are computationally intensive, and the presence of beam loss often prevents a smooth convergence toward the optimum. Under these circumstances, Bayesian optimization was adopted as a suitable approach to efficiently explore the parameter space and guide the optimization process.

Bayesian optimization is a sequential, model-based global optimization technique, where both the surrogate model and the decision-making process are updated after each function evaluation. As reviewed in Refs. [18] and [19], this technique is well established and widely applied in particle accelerator science and technology. In this optimization, the fixed beamline lattice is the part of the surrogate model. Since the underlying beam dynamics through the channel cannot be explicitly modeled in closed form, the beamline is treated as a black-box. To initialize the evolution, 120 data points are randomly selected from the six-dimensional variable space defined by the six input beam parameters within their respective limits, as given in Table 1. Therefore, these data points form the optimization variables, represented by a two-dimensional array with 120 rows, denoted by $x_i^m$, where $i=1,\ldots,6$ corresponds to the two transverse emittances $\varepsilon_x, \varepsilon_y$ and the corresponding Twiss parameters ($\alpha_x, \beta_x, \alpha_y, \beta_y$) and $m$ indicating the generation index. The column number is defined by the dimension of the reference data set. As mentioned earlier, a multiparticle tracking simulation involving fifty-thousand macro-particles is carried out using TRACEWIN for each data point within the population, and the corresponding transmission ratios are estimated from the simulation results. The objective function $f_1(x_i^m)$ is then formulated as the difference between the transmission ratios obtained from the simulations and those in the reference dataset. With advancement in the iteration, the optimization seeks the set of variables $x_i$ that minimizes the objective function $f_1(x_i^m)$. Optimization process then constructs a surrogate model using Gaussian Process (GP) to approximate $f_1(x_i^m)$. The GP provides predictions of both the mean $\mu(x_i^m)$ and the uncertainty $\sigma(x_i^m)$ for each data points in the population. The next set of variables, $x_i^{m+1}$, is selected based on these predictions, as guided by the acquisition function $A(x_i)$. The algorithm then balances exploration by testing regions of high uncertainty and exploitation by focusing on regions predicted to yield promising solutions through maximizing $A(x_i)$. The set of data points so chosen, is then used to evaluate the true objective function $f_1(x_{i,chosen})$, and the surrogate model is updated with this new result. The iteration is then repeated targeting the desired convergence.

The entrance emittances and Twiss parameters inferred in this first stage of the simulation-based semi-numerical framework are constrained by the measured transmission ratios across representative optics settings. However, transmission measurements alone cannot uniquely determine all beam distribution moments. Therefore, the goal is not a unique physical reconstruction but a model-consistent effective input beam parameter set that anchors the subsequent optics optimization within a realistic operational regime.

2.3 *Numerical Optimization of Quadrupole Magnet Settings*

The second part of this study focuses on the optimization of beamline optics with the objective of maximizing the beam transmission. The optimization is performed solely by adjusting the gradients of the quadrupole magnets for each beamline at a fixed energy. In this case, the beam parameters at the entrance of the transport line are known from the previously described Bayesian optimization. These beamlines, as mentioned, consist of six quadrupole magnets and the gradients of these quadrupoles are treated as the six optimization variables in the present study. The allowed range for these gradients is restricted to 0.25 to 5 T/m. The polarities of the quadrupole magnets are fixed and both beamline RTCBL-01 and RTCBL-04 have an identical quadrupole configurations- FQ, DQ, DQ, FQ, FQ, DQ. Here, FQ denotes a quadrupole focusing in the horizontal plane, while DQ denotes a quadrupole defocusing in the same plane.

Similar to the previous part, another Python-based optimization framework is developed here, with TRACEWIN invoked as a sub-process to perform beam dynamics simulations. Even though the input beam parameters are known, significant beam losses during transport, inherent to the beamline structure, make the problem analytically intractable. Therefore, similar to the previous case, this gradient optimization task also represents a nonlinear, higher-dimensional black-box problem. In view of this, the stochastic, derivative-free optimization algorithm Covariance Matrix Adaptation Evolution Strategy (CMA-ES) was implemented via a dedicated Python subroutine. Its performance has been demonstrated in recent studies [14] [20] through comparisons with other machine-learning-based optimization techniques. A comprehensive review of optimization methods in accelerator physics is available in Ref. [21].

In this formulation, beam transmission through each beamline is maximized by minimizing the objective function $\xi(y_j^n) = (1 - Ʈ^n)$, where $Ʈ^n$ denotes the transmission ratio recorded at the final Faraday cup upstream of the target and $y_j^n$ represents array with a number of the data points sampled within the gradient span. The number here refers to the population size, which is 60 in our case. The index $j = 1, \ldots .6$ corresponds to the gradients of the six quadrupoles magnets and $n$ is indicating the generation index. Offspring in each generation are produced by defining a lattice configuration, with quadrupole strengths chosen within their respective limits of 0.25–5 T/m. In each generation, offspring are sampled from the variable space using a multivariate normal distribution characterized by the mean vector $M^n$ and the global step-size $\sigma^n$ which evolves adaptively. The shape of the distribution is governed by the covariance matrix $\Lambda^n$, which defines the amplitude and orientation of the sampling distribution. Therefore, finally, the sampled offspring vector $y_j^n$ is expressed as $y_j^n \sim \mathcal{X}(M^{(g)}, [\sigma^{(g)}]^2 \Lambda^{(g)})$. According to CMA-ES algorithm, the optimization was initialized with a mean vector $M^0 = 0$ and a covariance matrix $\Lambda^0 = I$, representing an uncorrelated identity distribution. The global step-size $\sigma^0$ was set by the user to define the initial sampling scale, which was taken as 10 to 20% of the total variable span. Within the CMA-Evolution Strategy, population members are ranked in each generation based on their objective function values. The mean vector is updated as a weighted recombination of the top-ranked individuals. Here, the covariance matrix is adjusted to capture the statistical correlations among variables, enabling the distribution to align with the contours of the objective landscape. This iterative process continues until convergence criteria are met. Using the CMA-ES optimization, the optimized strength of the six quadrupole are determined to maximize beam transmission.

In this context, the suitability of Bayesian optimization and CMA-ES within the semi-numerical optimization framework is briefly discussed. Both Bayesian Optimization and CMA-ES are derivative-free global optimization techniques. Amongst, the Bayesian Optimization is employed to estimate the distribution parameters at the entrance sampled over a relatively wide domain, reflecting the lack of prior knowledge about their values and the inherently high uncertainty associated with them. Since it leverages a probabilistic surrogate model to balance exploration and exploitation while minimizing the number of expensive multiparticle simulations required, the Bayesian framework is well suited for this application. On the other hand, CMA-ES is used to maximize beam transmission using optimized the quadrupole magnet gradients. Compared to the first phase, the variables used here span a relatively narrower range and are better understood from prior operational experience, allowing for a more focused search. CMA-ES performs particularly well in such situations. It efficiently explores the local structure of the objective landscape. Its population-based sampling makes it robust to noise and multimodalities common in beam dynamics optimization.

We conclude this section with a brief summary of the developed simulation-based semi-numerical framework, followed by the workflow chart shown in Fig. 2. In this framework, the first stage estimates the unknown input beam parameters at the cyclotron exit using multiparticle simulation based Bayesian optimization. The experimentally measured transmission ratios for several operational lattice configurations provide the constraints for this semi-numerical estimation. In the second stage, a simulation-based CMA-ES optimization algorithm is used to maximize beam transmission through the beamline using the input beam parameters estimated in the first stage. In this implementation, the multiparticle tracking code TRACEWIN is used to support the Bayesian and CMA-ES optimization procedures. The two-stage semi-numerical optimization framework is executed offline. In the first stage, the input beam parameters are estimated, and in the second stage the optimized quadrupole gradients are obtained. The beam dynamics code TRACEWIN provides a high-fidelity multiparticle representation of nonlinear beam transport, including the effects of finite apertures and particle losses. However, small discrepancies between the simulation model and the real machine are expected. These differences may arise from factors such as magnet calibration uncertainties, alignment tolerances, and simplified field representations in the model. Finally, the near-optimal quadrupole settings obtained from the simulation-based optimizer are validated experimentally. Any remaining discrepancies are corrected through small operational trims within normal machine tolerances. The overall workflow of the developed semi-numerical optimization framework is summarized in the following schematic chart.

Stage-1: Estimate input beam parameters (cyclotron exit/beamline entrance)

---

    Prerequisites-   (a) orbit centering with steering magnets (Beam Viewer assisted),

                            (b) measured Transmission ratios at downstream Faraday Cups, and

                            (c) corresponding reference lattice config. (10 operational settings).

    Algorithm -      Bayesian Optimization.

    Simulation -     TRACEWIN multiparticle tracking.

    Variable space - 6D ($\varepsilon_x, \alpha_x, \beta_x, \varepsilon_y, \alpha_y, \beta_y$).

---

**Output: effective entrance beam parameters.**

↓

Stage-2: Quadrupole Gradient Optimization (RTCBL-01 and 04)

---

    Prerequisites-   (a) effective entrance Beam Parameters from Stage-1.

    Algorithm -      CMA-ES.

    Simulation -     TRACEWIN multiparticle tracking.

    Variable space - 6D (gradients of the six quadrupoles in the beamline lattice)

---

**Output: optimized quadrupole gradients that maximize simulated transmission at the last Faraday cup before target.**

↓

Stage-3: Experimental Verification and routine fine tuning (Beam line lattices)

---

    Prerequisites-   (a) orbit centering with steering magnets (Beam Viewer assisted),

                            (b) apply optimized Quadrupole settings

    Measurements – Beam transmission in the last Faraday Cup (FC-12 or FC-42)

---

**Acceptance: transmission performance plus operational constraints.**

**Fig**. 2: Workflow of the semi-numerical beamline optimization framework developed for transmission maximization in the RTC beamlines.

## 3. Results and Discussion

In this section, we present results of our semi-numerical optimization analysis described in the previous section. These results include the optimized beam parameters at the beamline entrance and the expected quadrupole settings corresponding to maximum beam transmission through the transport channels. Along with this, the comparison between the simulated and experimentally measured beam transmissions is also discussed. In this context, it may be noted that for cyclotrons such as the K-130 RTC at VECC, the extracted beam exhibits a finite energy spread arising from RF phase width, space-charge effects, and extraction dynamics. Typically, the energy spread lies in the range of about 0.2-1%, depending on the machine parameters and beam intensity [22-23]. Values around 0.3-0.6% are commonly achieved under optimized operating conditions, consistent with reported measurements of cyclotron beams [24]. In the present beam dynamics calculations, an energy spread of 0.6% has therefore been assumed to account for possible beam losses arising from dispersion in the dipole magnets.

*3.1 Estimation of input beam parameters*

In this subsection, we present the results of our semi-numerical analysis aimed at describing the input beam parameters at the entrance of each beamline corresponding to specific beam energies for the two beamlines RTCBL-01and RTCBL-04. As explained earlier, the dc beam at the entrance is modelled as a four-dimensional Gaussian distribution in the transverse phase space, sampled using fifty thousand macroparticles. The distribution is generated in TRACEWIN using its built-in distribution generator at the cyclotron exit, which also defines the entrance distribution for the transport line and serves as the initial condition for the subsequent beam transport analysis.

As mentioned earlier, the first phase of the computation is carried out for RTCBL-01 operating with 7 MeV and 13 MeV proton beams and for RTCBL-04 with an 11 MeV proton beam. In this phase, a Bayesian-optimization-based algorithm iteratively evolves a population of 120 individuals. The population is sampled across the six variables defining the four-dimensional Gaussian distribution of the input beam, with their respective ranges listed in Table 1. For each species, TRACEWIN-based multiparticle tracking embedded in the algorithm is performed for a few reference beamline lattice configurations. The corresponding transmission ratios are calculated at the downstream Faraday cups. These simulated transmission ratios are then compared with the experimentally measured values. The algorithm iteratively adjusts the six beam parameters defining the input distribution to minimize the difference between the simulated and measured transmission ratios. The converged results for the 7 MeV proton beam in RTCBL-01 under the reference lattice configurations are shown in Fig. 3. The corresponding optimum input beam parameters defining the four-dimensional Gaussian distribution at the beamline entrance are presented in Table 2.

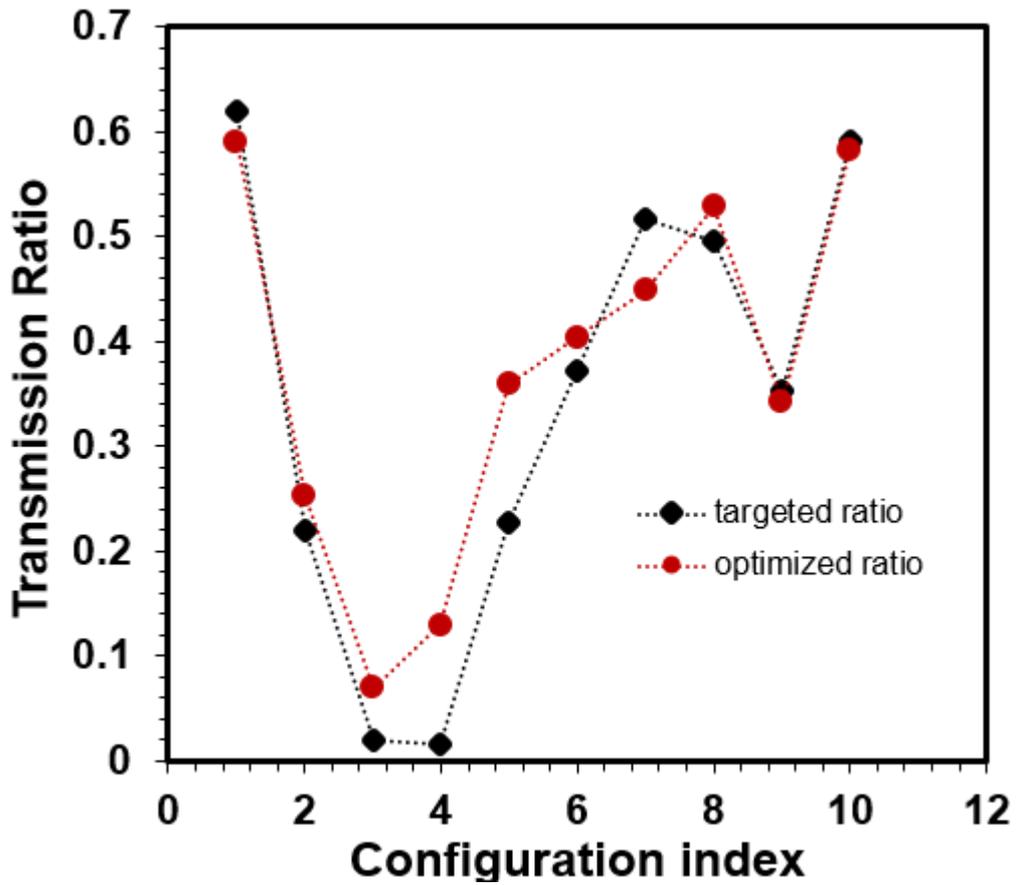

**Fig. 3:** Computed and experimentally obtained transmission ratios for different optics in RTCBL-01 for 7MeV proton beam.

**Table 2:** Computed beam distribution at the entrance of RTCBL-01 for 7MeV proton beam.

| $\varepsilon_x$ (mm mrad) | $\alpha_x$ | $\beta_x$ (mm/mrad) | $\varepsilon_y$ (mm mrad) | $\alpha_y$ | $\beta_y$ (mm/mrad) |
|---|---|---|---|---|---|
| 4.38 | -18.36 | 34.94 | 1.08 | -0.01 | 7.36 |

Figure 4 shows the TRACEWIN-generated beam distribution at the entrance of RTCBL-01 for a 7 MeV proton beam.

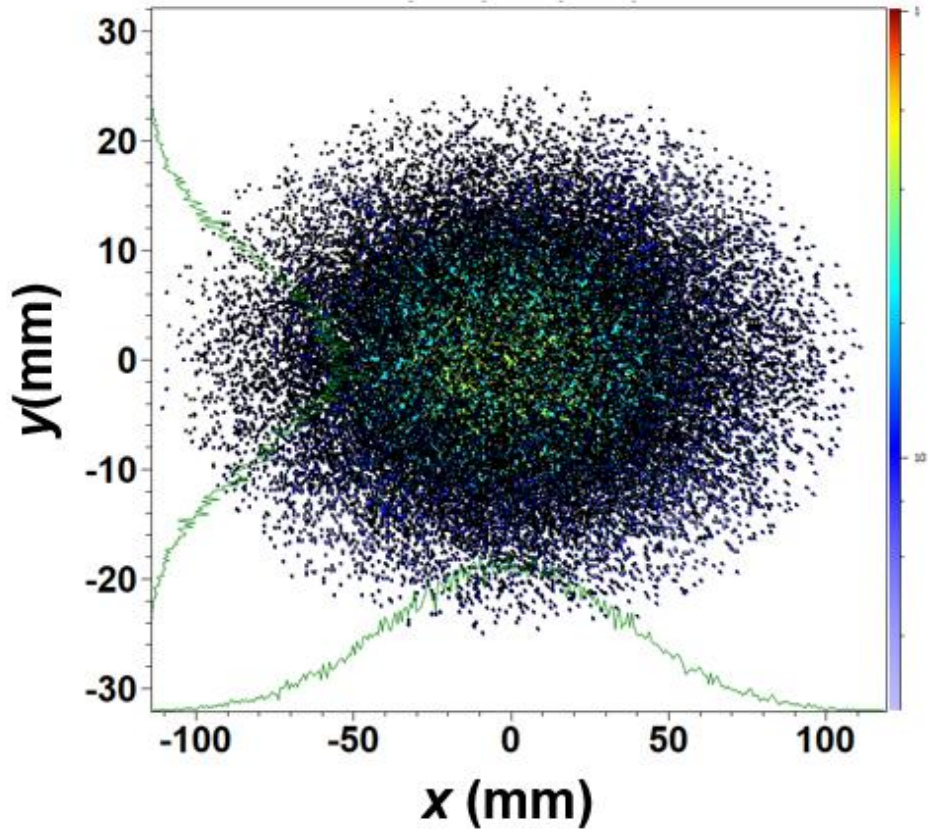

**Fig. 4:** TRACEWIN-generated input beam distribution at the cyclotron exit for RTCBL-01 at 7 MeV.

Since the transport line has a finite aperture of 100 mm, the transmitted beam corresponds to a truncated Gaussian distribution. Only a fraction of the particles from the initial distribution can pass through the entrance aperture, while the rest are lost at that location. The resulting effects on beam emittance and transmission are examined in the next subsection through a representative case study.

RTCBL-01 is further analyzed at proton beam energy of 13 MeV. Figure 5 shows a comparison between the experimentally measured transmission ratios and the simulation results under the reference lattice configurations. The optimized parameters used to generate the initial beam distribution at the entrance of the beamline are summarized in Table 3.

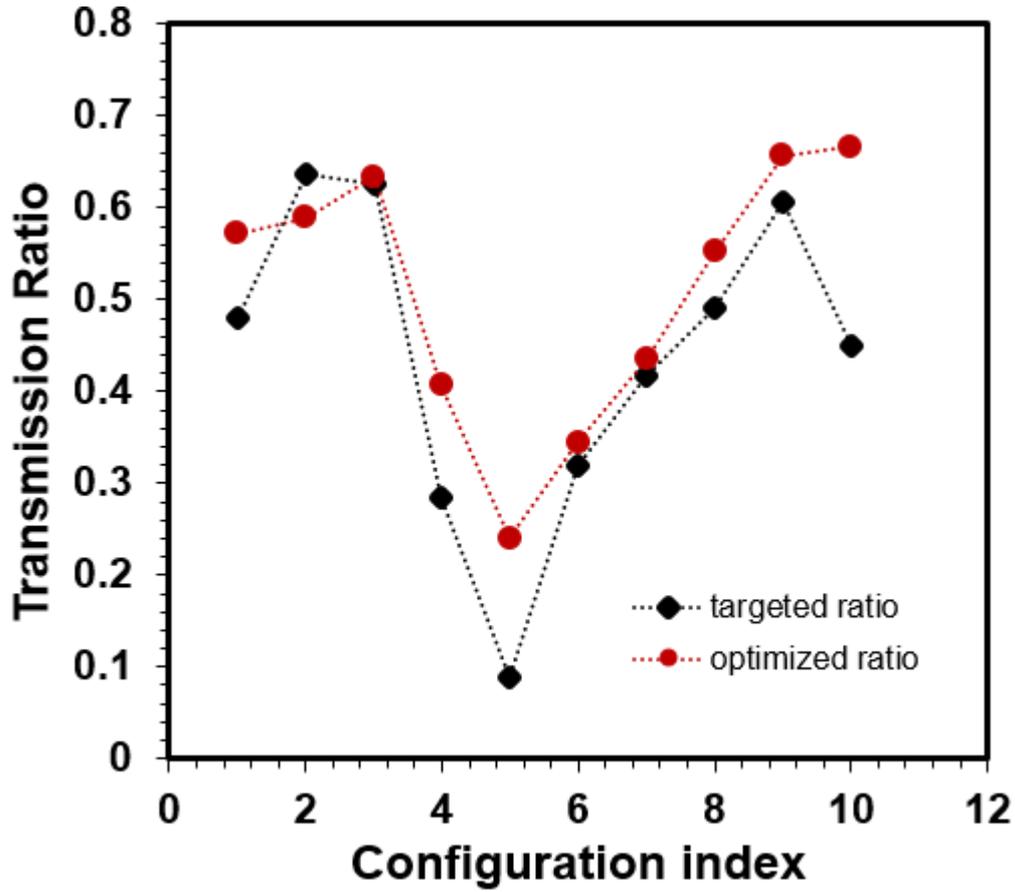

**Fig. 5:** Computed and experimentally obtained transmission ratios for different optics in RTCBL-01 at 13MeV.

**Table 3:** Optimized parameters used to generate the beam distribution at RTCBL-01 entrance at 13 MeV.

| $\varepsilon_x$ (mm mrad) | $\alpha_x$ | $\beta_x$ (mm/mrad) | $\varepsilon_y$ (mm mrad) | $\alpha_y$ | $\beta_y$ (mm/mrad) |
|---|---|---|---|---|---|
| 5.00 | -23.3 | 43.76 | 1.62 | -0.01 | 10.54 |

A similar analysis was performed for the RIB beamline RTCBL-04. An energy of 11 MeV is selected in this case to serve as an illustrative example. The comparison between the measured and simulated transmission ratios under the reference lattice configurations is shown in Fig. 6. The optimized initial beam distribution parameters for this case are summarized in Table 4.

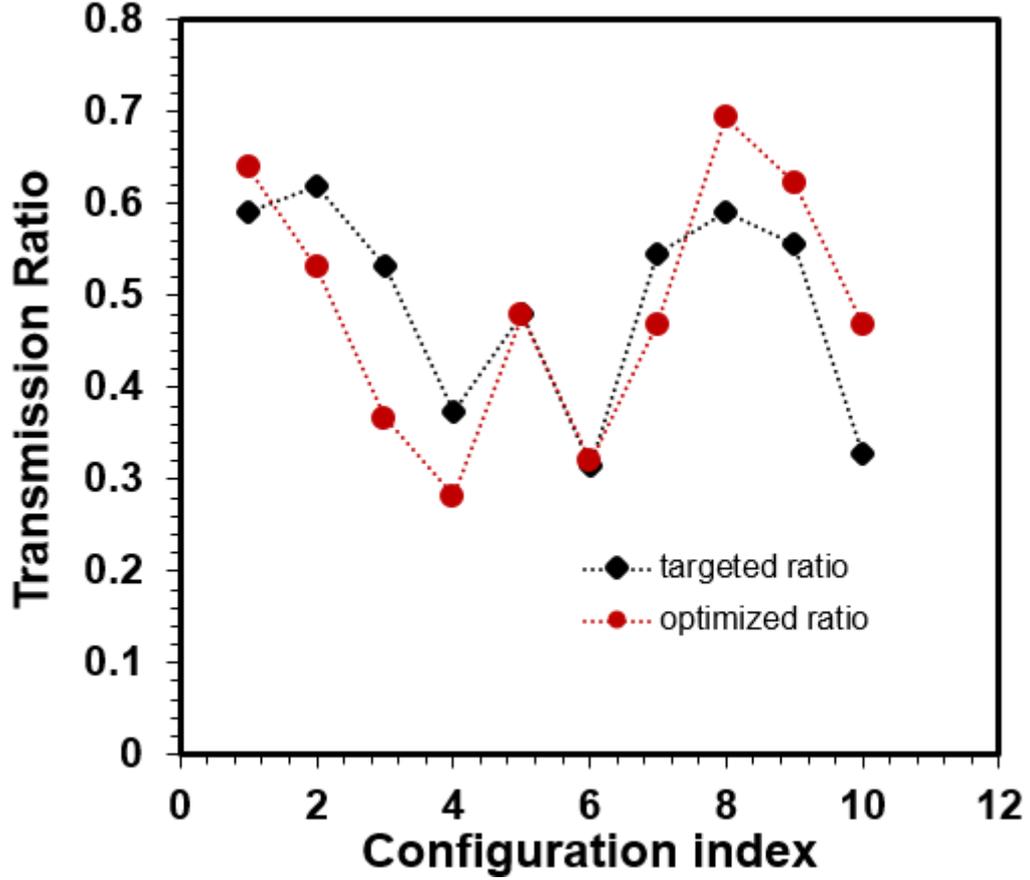

**Fig. 6:** Computed and experimentally obtained transmission ratios for different optics in RTCBL-04 at 11 MeV.

**Table 4:** Optimized parameters used to generate the beam distribution at RTCBL-04 entrance at 11 MeV.

| $\varepsilon_x$ (mm mrad) | $\alpha_x$ | $\beta_x$ (mm/mrad) | $\varepsilon_y$ (mm mrad) | $\alpha_y$ | $\beta_y$ (mm/mrad) |
|---|---|---|---|---|---|
| 5.84 | -13.61 | 24.92 | 0.68 | -1.39 | 5.37 |

Besides beam current, experiments often require specific beam profiles or spot sizes at the target, realized through different optics configurations in the K-130 RTC at VECC. Each optics configuration corresponds to a dataset of beam transmission measurements obtained at the downstream Faraday cups. In this work, we use a reference dataset consisting of 10 operational lattice configurations selected from these commonly used optics settings for the beamline at a given beam energy. These configurations represent typical operating conditions used during routine beamline tuning and therefore provide representative transmission responses of the beamline lattice. These reference datasets are used only as benchmark constraints to guide the Bayesian optimization toward a realistic region of the input beam parameter space. However, as mentioned earlier, the optimization itself explores a much larger parameter domain by evolving a population of 120 candidate solutions in the six-dimensional variable space. Each candidate solution is evaluated through multiparticle tracking simulations involving a large number of macroparticles. Thus, the reference dataset does not restrict the search space but serves to guide the estimated input parameters toward consistency with experimental observations. In practice, the optimized beam parameters reproduce the measured transmission ratios across the reference configurations with reasonable agreement, as shown in Figs. 3, 5, and 6. Therefore, although the reference dataset contains only 10 configurations, it provides sufficient constraints to reliably determine realistic input beam parameters within the present optimization framework.

*3.2 Computation of Optimized Quadrupole Parameters*

In the preceding section, we presented an estimation of the optimized beam conditions at the entrance of the respective beamlines. The optimized quadrupole magnet settings for RTCBL-01 and RTCBL-04 at their respective target energies are presented here, with the objective of achieving maximum beam transmission at the last Faraday Cup just before the target. As described in Section 2.3, the CMA-ES–based algorithm is applied to optimize the quadrupole gradients in the respective beamline lattices at different energies, using the input beam conditions described in the previous subsection.

Table 5 presents the optimized quadrupole gradient values computed using the TRACEWIN-embedded CMA-ES–based algorithm. The calculation uses the optimized input beam distribution parameters listed in Table 2 for RTCBL-01 at a beam energy of 7 MeV. Multiparticle beam tracking with the optimized configuration predicts a beam transmission of approximately 76% at the location of FC-12.

To verify this experimentally, the optimized quadrupole gradient values obtained from the CMA-ES–based semi-numerical framework were applied to the six quadrupole magnets of RTCBL-01. As discussed earlier, orbit centring using steering magnets is treated as a prerequisite in this framework. In addition, small trims of the optimized quadrupole set-points were finally applied as 'minor-adjustments' during beam tuning to satisfy both transmission and beam spot-quality requirements at the destination.

The corresponding experimental quadrupole settings and measured beam transmission are listed in the last row of Table 5. Using these experimentally implemented settings, a beam transmission of approximately 73% was measured at FC-12, which is labelled as the 'Verified' data. These transmission values are calculated from the measured beam currents obtained using identical conventional water-cooled intercepting Faraday cups equipped with secondary-electron suppression.

**Table 5:** Computed vs. Verified Quadrupole (QM) Gradients (in T/m) and Transmission in RTCBL-01 at 7 MeV

| Case | QM-1 | QM-2 | QM-3 | QM-4 | QM-5 | QM-6 | Transmission |
|---|---|---|---|---|---|---|---|
| computed | 2.57 | -2.40 | -2.10 | 2.02 | 2.56 | -3.19 | ~76% |
| verified | 2.58 | -2.28 | -2.40 | 2.14 | 2.54 | -3.26 | ~73% |

Before presenting the comparison between the calculated and experimentally verified data for the other two cases (the 11 MeV case in RTCBL-04 and the 13 MeV case in RTCBL-01), we briefly discuss the characteristics of the beam transmitted through RTCBL-01 at 7 MeV. Figure 7 shows a representative example of the evolution of the transverse beam envelopes along the beam line for the optimized optics configuration listed in Table 5. The red and blue curves represent the horizontal and vertical *rms* beam sizes, respectively.

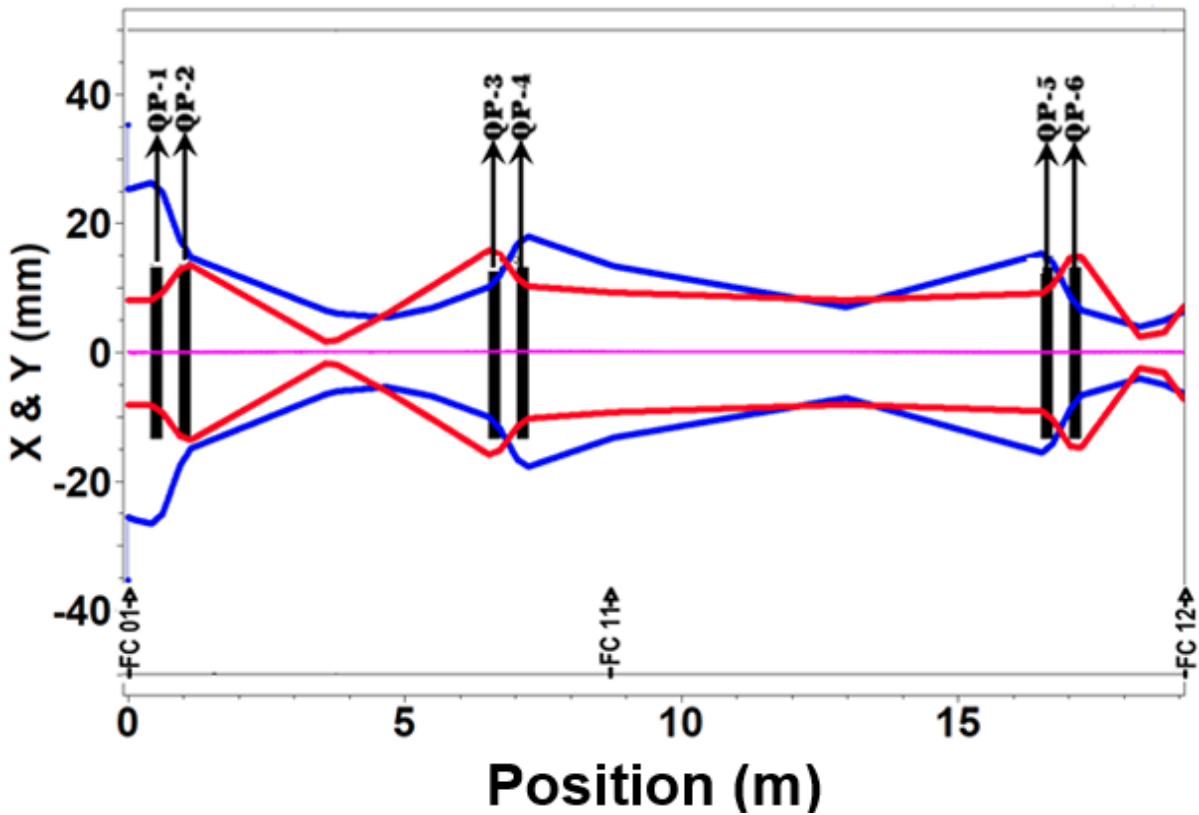

**Fig. 7:** Beam envelope evolution along the RTCBL-01 beamline for 7 MeV protons using the optics configuration listed in Table 5. Red and blue curves denote the horizontal (x) and vertical (y) *rms* beam envelopes, respectively. Black blocks indicate the locations of the six quadrupole magnets (QP1–QP6). FC01, FC11, and FC12 denote the three Faraday cup positions.

Figure 8 shows the evolution of the two transverse *rms* beam emittances along RTCBL-01 for the above-mentioned optics configuration. Among the two, the horizontal *rms* beam emittance exhibits a noticeable drop near the entrance region. This behaviour originates from the assumed beam distribution at the cyclotron exit. As described earlier, the initial beam is modelled as a four-dimensional Gaussian distribution represented by 50,000 macroparticles generated in TRACEWIN, spanning the transverse phase space up to 3 standard deviations (corresponding to a ratio of 3 between the maximum and *rms* beam size). Consequently, a significant fraction of particles lies outside the physical acceptance of the 100 mm diameter transport line and is therefore lost at the entrance plane, effectively truncating the initial Gaussian distribution.

This truncation is also reflected in the beam transmission shown in Fig. 9. According to the simulation, the beam transmission at the first Faraday cup corresponds to about 48% of the initial macroparticles. Such phase-space truncation is expected when the beam extracted from the electrostatic deflector channel enters the limited acceptance of the external beamline aperture in compact cyclotrons such as the K-130 RTC at VECC, Kolkata. The initial decrease in the horizontal *rms* beam emittance therefore reflects successive particle losses occurring during the first few tracking steps after the beam enters the transport line.

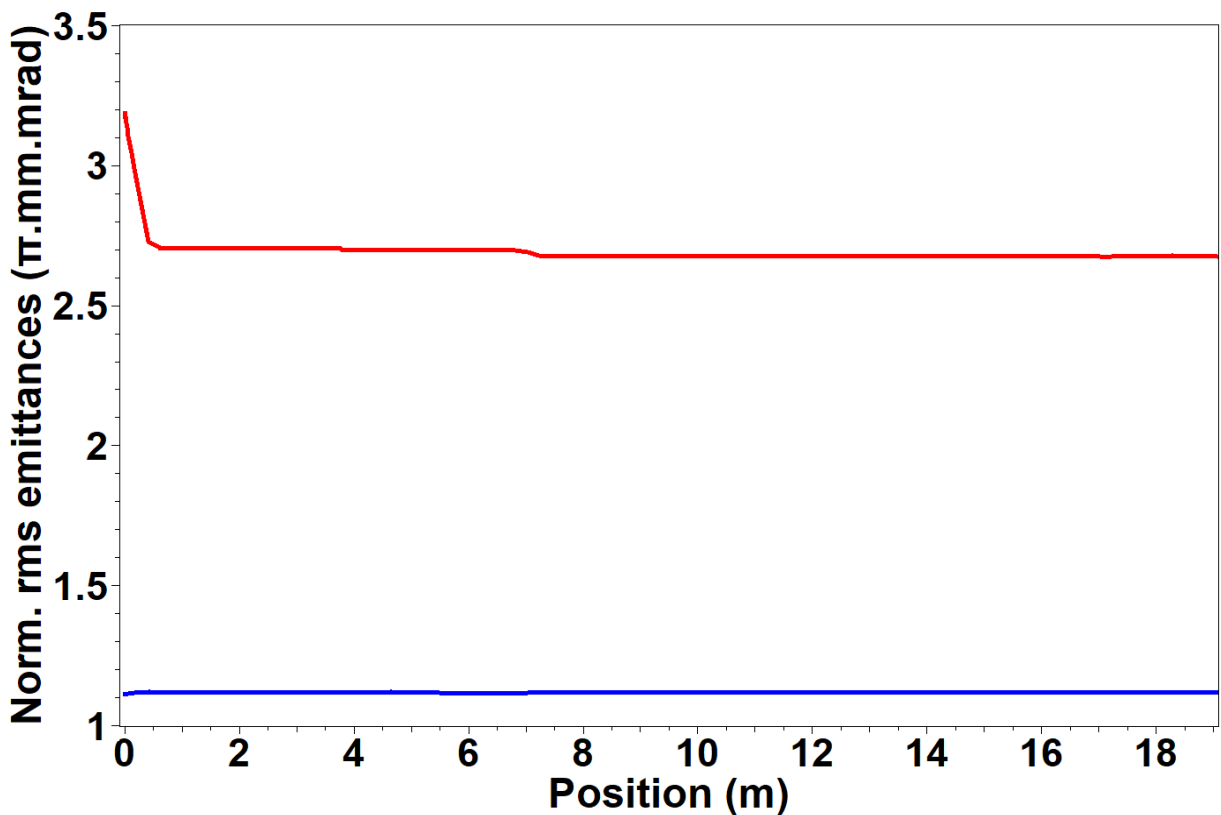

**Fig. 8** Evolution of the normalized *rms* emittances along RTCBL-01. The red curve corresponds to the horizontal (*x*) emittance and the blue curve to the vertical (*y*) emittance.

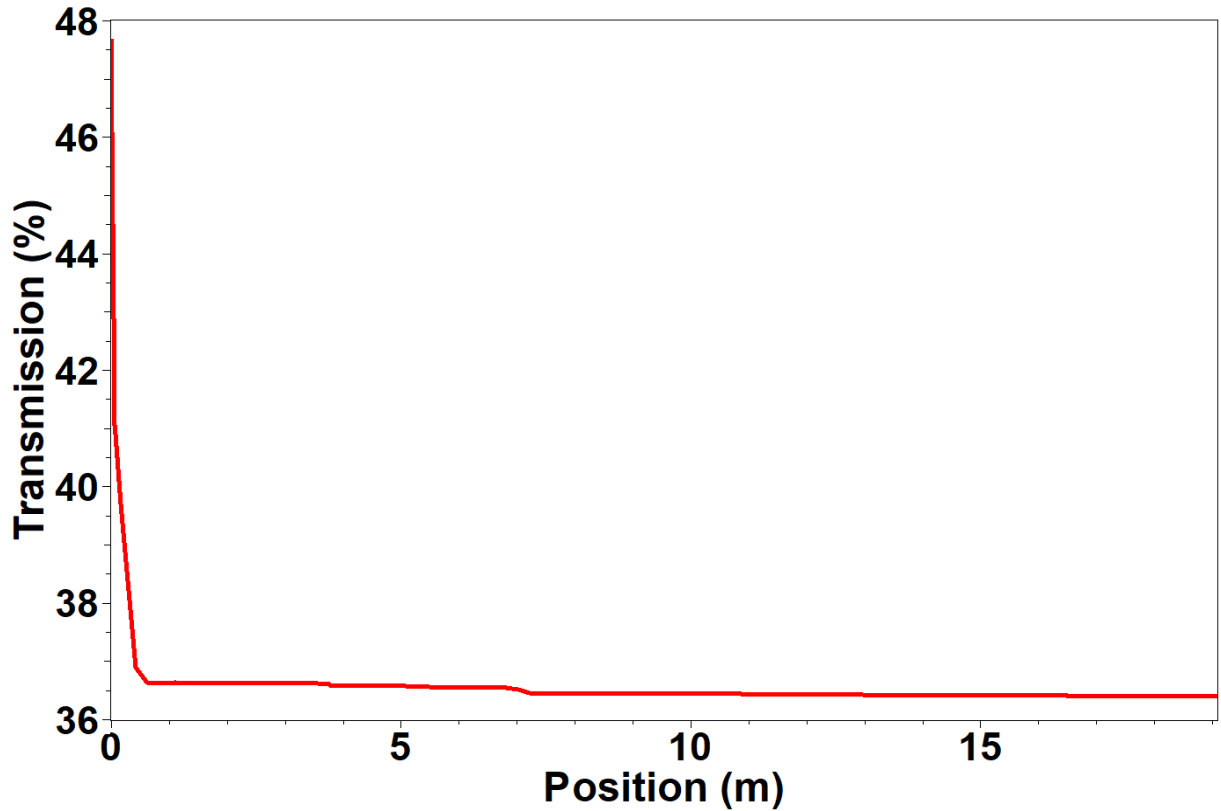

**Fig. 9:** Beam transmission (%) along the transport line.

We now present the remaining analysis comparing the calculated and experimentally verified data for the 13 MeV beam in RTCBL-01 and the 11 MeV beam in RTCBL-04. The beamline was optimized for a 13 MeV proton beam using the quadrupole settings listed in Table 6 to maximize the transmission at FC-12. Multiparticle tracking predicts a transmission of approximately 74%. Applying these settings to RTCBL-01 with minor operational tuning yielded an experimental transmission of about 69% at FC-12.

**Table 6:** Computed vs. Verified Quadrupole (QM) Gradients (in T/m) and Transmission in RTCBL-01 at 13 MeV

| Case | QM-1 | QM-2 | QM-3 | QM-4 | QM-5 | QM-6 | Transmission |
|---|---|---|---|---|---|---|---|
| computed | 3.59 | -3.23 | -2.79 | 2.27 | 2.93 | -3.74 | ~74% |
| verified | 3.48 | -3.11 | -3.04 | 3.00 | 3.11 | -3.59 | ~69% |

Similar analysis is also performed for the RIB beamline RTCBL-04 with an 11 MeV proton beam. As illustrated in the Fig. 2, the beam line includes a 594 mm long analyzing magnet that bends the transported beam by $160^0$. This beam transport line is analyzed at a beam energy of 11 MeV. The quadrupole gradient values obtained from multiparticle tracking, as well as those applied in the RTC beamline corresponding to approximately 81% and 77% beam transmission, are presented in Table 7. At this point, we make an important observation regarding the gradients of QM-5 and QM-6 of RTCBL-04. The verified gradients for these magnets are about ±1.03 T/m, which are significantly smaller in magnitude than the computed optimal values. This is because the experimental tuning was performed not only to maximize the beam current at the final Faraday cup (FC-12) but also to obtain a uniform beam spot at the target station. The verified values represent the lower bounds of the absolute gradient range that still provides maximum transmission. Within this range, the beam current remains nearly constant, while the beam profile and uniformity vary. For QM-5, this range extends approximately from 1.03 to 1.78 T/m, and for QM-6 from −1.03 to −1.62 T/m.

Table 7: Computed vs. Verified Quadrupole (QM) Gradients (in T/m) and Transmission in RTCBL-04 at 11 MeV

| Case | QM-1 | QM-2 | QM-3 | QM-4 | QM-5 | QM-6 | Transmission |
|---|---|---|---|---|---|---|---|
| computed | 3.28 | -3.01 | -2.95 | 2.42 | 1.50 | -1.50 | ~81% |
| verified | 3.21 | -2.88 | -3.29 | 2.85 | 1.03 | -1.03 | ~77% |

## Conclusion

The demand for high beam current at the experimental stations remains a continual requirement from the K130 user community at VECC. However, the non-availability of measured beam parameters at the beamline entrances makes it difficult to optimize the quadrupole gradients for maximum transmission in this decades-old machine.

To address this challenge, a Python-based semi-numerical optimization framework integrated with TRACEWIN has been developed and presented here to determine suitable lattice configurations for maximizing beam transmission in this room-temperature cyclotron at VECC. In the first phase, the developed framework estimates the input beam parameters through a Bayesian optimization based approach. The beam line lattice is optimized in the next phase using the CMA-Evolution Strategy to maximize beam transmission at the targeted location. Using these optimized settings, along with minor adjustments, the measured beam transmission has exceeded 70% in most cases, marking a notable improvement over the typical 60% transmission usually observed in the control room. These experimentally measured maximum beam transmissions are in good agreement with the values obtained from our optimization analysis. In this context, it should be noted that beamline tuning in the RTC at VECC is performed not only to maximize the beam current delivered to the target. In addition, considerable emphasis is placed on meeting user-specified beam quality requirements, such as the desired beam spot size and beam uniformity at the target. From this overall operational perspective, the improvement in beam transmission to values exceeding 70% reported in this work represents a reasonable and meaningful enhancement.

The developed two-stage semi-numerical optimization framework is general in nature. The first stage determines an effective set of input beam parameters without the need for additional hardware or interruption to routine machine operation. In the second stage, these parameters are used to optimize the beam transport line. As a result, the methodology can be implemented without disrupting normal machine operation.

In this respect, the underlying philosophy of the present simulation-based semi-numerical approach is not to reconstruct the exact input beam parameters in full detail. Instead, it aims to obtain a set of representative parameters that can reliably reproduce the observed beam transport behaviour. This distinguishes the present approach from conventional diagnostic methods, such as quadrupole-scan techniques. Such methods are designed to determine the beam parameters with high accuracy through dedicated measurements. We therefore conclude this paper with a brief comparison between the machine-specific operational optimization framework developed in this work and the well-known study by E. Prat et al. on four-dimensional transverse beam matrix measurement using the quadrupole scan technique [25]. Their multiple-quadrupole scan method provides a direct and compact approach for reconstructing the ten independent second moments of the four-dimensional transverse beam matrix, and therefore offers rich diagnostic information, including the explicit identification of cross-plane coupling. In contrast, the present work does not aim to perform a full 4D beam-matrix measurement. Instead, it addresses the practical problem of optimizing beam transmission in the existing K130 RTC extraction lines, where the installation of additional diagnostics and dedicated quadrupole-scan measurements is difficult.

In this sense, the machine-specific operational optimization framework developed here represents a distinct class of method, designed primarily for practical deployment in constrained legacy accelerator systems.

**Acknowledgments:**

The authors gratefully acknowledge Dr. Vinit Kumar and Dr. S. Yadav of RRCAT, Indore, for their guidance and support in evolutionary optimization methods. They are deeply grateful to Dr. Arup Bandyopadhyay of VECC, Kolkata, for his constant encouragement and support, and to the RTC control room staff for their valuable assistance. The authors also acknowledge the unknown reviewers of this paper for the critical assessment and evolution of the work and their worthy suggestions.